# Rust-accelerated powder X-ray diffraction simulation for high-throughput and machine-learning-driven materials science

Miroslav Lebeda[a,b,c], Jan Drahokoupil[a,c], Petr Veřtát[c], Petr Vlčák[a]

[a] *Faculty of Mechanical Engineering, Czech Technical University in Prague, Technická 4, 16607 Prague 6, Czech Republic*

[b] *Faculty of Nuclear Sciences and Physical Engineering, Czech Technical University in Prague, Trojanova 339/13, 12000 Prague 2, Czech Republic*

[c] *FZU – Institute of Physics of the Czech Academy of Sciences, Na Slovance 2, 18200 Prague 8, Czech Republic*

**Corresponding author:** Miroslav Lebeda, lebedmi2@cvut.cz

## Abstract

High-throughput powder X-ray diffraction (XRD) simulations are a key prerequisite for generating large datasets used in the development of machine-learning models for XRD-based materials analysis. However, the widely used pymatgen powder XRD calculator, implemented entirely in Python, can be computationally inefficient for large-scale workloads, limiting throughput. We present XRD-Rust, a Rust-accelerated implementation of the pymatgen powder XRD calculator that maintains compatibility with existing Python-based workflows. The method retains pymatgen for crystal structure handling while reimplementing the computationally intensive parts of the XRD calculation in Rust, with optional further acceleration via SIMD vectorization and multi-threaded execution across reflections using the Rayon library. Performance benchmarking on two large crystallographic datasets, the Materials Cloud Three-Dimensional Structure Database (MC3D, 33 142 structures) and the Crystallography Open Database (COD, 515 181 structures), demonstrates substantial speedups. For MC3D, XRD-Rust achieves a median serial SIMD speedup of 10.8× (median absolute deviation, MAD, 1.8×), increasing to 15.1× (MAD 3.8×) with 8-thread parallel execution, with a maximum runtime reduction from 40.5 s to 0.9 s. For COD, the median serial SIMD acceleration reaches 10.7× (MAD 4.2×), while 8-thread parallel execution yields 19.5× (MAD 10.0×), with a maximum reduction from 1437 min to 1 min. SIMD vectorization alone provides additional performance gains ranging from a few percent to several tens of percent, depending on the workload and execution mode. Overall, these benchmarks demonstrate that XRD-Rust significantly accelerates powder XRD simulations compared to the original pymatgen implementation, enabling efficient high-throughput dataset generation and improving performance in interactive diffraction analysis applications.

**Keywords:** Powder X-ray diffraction (XRD), Rust, Python, Pymatgen, High-throughput materials characterization

## 1. Introduction

Powder X-ray diffraction (XRD) is one of the most widely used techniques for materials characterization. Computational simulation of powder XRD patterns from known crystal structures plays a central role in phase identification, structure validation, and construction of reference databases. In recent years, simulated diffraction patterns have also become an essential component of data-driven materials science, particularly for training and validating machine-learning (ML) models for automated XRD analysis[1].

A growing number of works have demonstrated that ML models can perform XRD-related tasks such as crystal system classification[2–4], phase identification[5,6], qualitative phase analysis[7], and ML-assisted Rietveld refinement[8]. These models typically rely on very large synthetic training datasets generated from experimentally known or theoretically predicted crystal structures, often comprising millions of diffraction patterns modified with experimental effects such as peak broadening, noise, and preferred orientation. Similar computational demands arise in generative models for crystal structure prediction[9] and in active-learning experimental workflows[10], where diffraction patterns must be calculated repeatedly during optimization. The crystal structures used to generate such datasets are commonly obtained from large crystallographic databases, for instance, from: Materials Project (MP, publicly available)[11], Crystallography Open Database (COD, publicly available)[12], Open Quantum Materials Database (OQMD, publicly available)[13], Materials Cloud Three-Dimensional Structure Database (MC3D, publicly available)[14], or Inorganic Crystal Structure Database (ICSD, licenced)[15].

The generation of such large datasets imposes demands on the efficiency of powder XRD simulation. Within widely employed Python-based workflows, an established tool is the XRD calculator implemented in the pymatgen library (MIT license)[16]. This implementation incorporates kinematical diffraction theory, atomic scattering factors, Lorentz-polarization (LP) corrections, and optional thermal effects. Due to its availability in the broad pymatgen Python package, it can be easily employed in the workflow for high-throughput calculations of diffraction patterns. As a result, it has become one of the standards for Python-based crystallographic analysis. However, since pymatgen's XRD calculator is implemented in pure Python, its performance can become a significant bottleneck in high-throughput and ML-oriented workflows, particularly for large unit cells, low-symmetry structures, or the batch processing of a large number of structures.

Besides Python-based pymatgen, several established crystallographic software packages provide functionality for simulating powder X-ray diffraction patterns. These include publicly available tools such as Profex[17], GSAS-II[18], Maud[19], Jana2020[20], as well as licensed commercial software including Rigaku Smartlab Studio, PANalytical HighScore[21], and TOPAS[22]. Visualization-oriented tools such as publicly available VESTA[23] or the licensed software Diamond[24] are also commonly employed. While these packages are widely used for diffraction analysis and simulation, they are primarily designed for interactive or file-based workflows and offer limited integration with Python-based data pipelines and machine-learning frameworks.

Additionally, the closest alternatives to the pymatgen's XRD calculator among pythonic implementations appear to be Dans Diffraction (github.com/DanPorter/Dans_Diffraction) and Xrayutilities[25]. Dans Diffraction seems more suitable for smaller calculations rather than large-scale workflows due to its printing of information into the console and being fully Pythonic like pymatgen. On the other hand, Xrayutilities is a Python package with several routines written in C, offering a significantly faster alternative to pymatgen. However, since pymatgen is currently one of the most widely employed packages in materials science, providing a faster alternative directly compatible with pymatgen notation is highly beneficial for already implemented workflows or for researchers already familiar with it.

As an example, in XRDlicious (xrdlicious.com)[26], our recently developed browser-based interactive application for powder diffraction calculation that employs pymatgen, a noticeable performance reduction is observed when processing for instance large organic crystal structures containing over hundreds of atoms, particularly when diffraction patterns from multiple structures are uploaded simultaneously for comparison. In such cases, the time required to generate diffraction patterns can increase to tens of seconds or more, significantly impairing the user experience. This further confirms the need for more optimized implementations that combine the performance, usability and interoperability of Python-based workflows while notably reducing computation times.

To address these limitations, we present XRD-Rust, a Rust-accelerated implementation of the existing pymatgen's powder XRD calculator. The approach does not introduce new diffraction algorithms but focuses on implementation-level performance optimization while maintaining full compatibility with existing pymatgen-based workflows. Rust is a compiled programming language designed for high-performance execution of computationally intensive tasks[27,28]. Our approach retains pymatgen's functionality for crystal structure handling, while accelerating the performance-critical components of the calculation using Rust. The performance of the Rust-accelerated implementation is benchmarked on two large crystal structure databases, MC3D and COD.

For the MC3D dataset, a serial Rust-accelerated calculator with SIMD vectorization achieves a median speedup of 10.5× with a median absolute deviation (MAD) of 1.8×. The maximum calculation time reduction was from 40.5 s to 0.9 s. For the COD dataset, the corresponding median speedup is 10.7× (MAD 4.2×) and the maximum time reduction is from 1437 min to 1 min. Parallelization with the Rayon library on 8 threads with SIMD further improves the median speedup to 14.6× (MAD 3.8×) for MC3D and to 19.5× (MAD 10.0×) for COD.

While XRD-Rust is not intended to replace established, highly optimized crystallographic programs implemented in compiled languages such as C, C++, or Fortran, it significantly improves the computational efficiency of powder XRD simulations within the Python ecosystem, particularly in workflows based on pymatgen. By combining the flexibility of Python and its integration with machine-learning frameworks with Rust-based acceleration, the proposed approach occupies a complementary position between ease of use and high performance. Ultimately, this enables scalable, high-throughput simulations while preserving seamless compatibility with existing pymatgen workflows

## 2. Implementation Details

The source code and its documentation are publicly available on GitHub at github.com/bracerino/xrd-rust and can be installed from the Python Package Index (PyPI) (pypi.org/project/xrd-rust) using '*pip install xrd-rust*'. The package is distributed for Linux, macOS, and Windows platforms.

Rather than completely rewriting pymatgen's XRD calculator, we adopted a targeted optimization strategy that combines Python and Rust. The Python pymatgen layer is used for crystallographic operations with low computational cost, including parsing crystal structure data and applying space-group symmetry. The performance-critical numerical components were reimplemented in Rust. These include the generation of reciprocal lattice points within the physically accessible sphere, the nested loops for structure-factor evaluation by summing atomic contributions for each reflection, pairwise operations for grouping symmetry-equivalent reflections, the application of geometric and physical correction factors (Lorentz-polarization and Debye-Waller), and the merging of reflections within specified angular tolerances. Additionally, multi-threading across reflections is provided via the Rayon data-parallelism library (github.com/rayon-rs/rayon), with the number of threads configurable by the user. The structure-factor inner loop is further accelerated using SIMD vectorization via the wide package (github.com/Lokathor/wide).

The Rust-accelerated implementation does not apply any built-in peak filtering and instead returns the complete list of calculated reflections. Optional filtering can be performed subsequently as a post-processing step if desired. By contrast, the original pymatgen implementation automatically removes reflections with intensities below 0.1% of the maximum peak intensity. This behaviour is not exposed as a user-configurable parameter and can only be modified by modifying the pymatgen package source code. Because this filtering step is computationally inexpensive (tens of thousands of reflections typically requiring only a few milliseconds to determine it), it does not affect performance comparisons between the two approaches.

The Rust-accelerated implementation interfaces with Python through PyO3 (github.com/PyO3/pyo3), which provides bindings between Rust and the Python. The Rust code is compiled into a platform-specific shared library using Maturin (github.com/PyO3/maturin), a build system designed for packaging Rust-Python extensions. The compiled libraries can be imported like standard Python modules, allowing Rust functions to be called directly from Python while PyO3 automatically handles data conversion between the two languages. The interface accepts standard pymatgen structure objects and returns diffraction patterns in the same format as the original pymatgen implementation, requiring no major modifications to existing workflows. A minimal example script demonstrating powder XRD pattern calculation with the Rust-accelerated package is shown in **Fig. 1**.

```python
from pymatgen.core import Structure
from xrd_rust_calculator import XRDCalculatorRust

# Load structure and calculate powder XRD pattern
structure = Structure.from_file("structure.cif")
calc = XRDCalculatorRust(wavelength="CuKa", symprec = 0,
    parallel = True, num_threads  = 4, use_simd = True)
pattern = calc.get_pattern(structure, scaled=False, two_theta_range=(5, 70))

# Save to file
with open("xrd_pattern.csv", 'w') as f:
    f.write("2theta,intensity,hkl,multiplicity\n")
    for i in range(len(pattern.x)):
        hkl_list = [tuple(h['hkl']) for h in pattern.hkls[i]]
        hkl_str  = str(hkl_list)
        mult     = sum(h['multiplicity'] for h in pattern.hkls[i])
        f.write(f"{pattern.x[i]},{pattern.y[i]},{hkl_str},{mult}\n")
```

**Fig. 1:** Minimal Python script demonstrating powder XRD pattern calculation using the Rust-accelerated calculator with 4-thread parallelization via the Rayon library and SIMD vectorization.

## 3. Numerical Validation

To validate the numerical consistency of XRD-Rust with the original pymatgen implementation, we compared peak positions and intensities up to eight decimal places across the first 1 000 sorted structures from the COD containing between 500 and 1 500 atoms (**Fig. 2**). For each structure, the root mean square error (RMSE) was computed, followed by an overall RMSE across all structures, comprising a total of 8 638 808 reflections. Peak positions calculated by XRD-Rust show excellent agreement, with a mean RMSE of 0.00001552° (maximum RMSE of 0.00058562°), well below typical experimental uncertainty. Peak intensities, normalized to a 0 – 100 scale, exhibit a mean RMSE of 0.000614% relative to the maximum peak intensity. No reflection showed an intensity deviation exceeding 1% of the maximum intensity, and the worst-case RMSE across all structures remained as low as ~0.023 %. These results demonstrate that XRD-Rust reproduces pymatgen outputs with high numerical fidelity, with no physically meaningful deviations in peak positions or intensities.

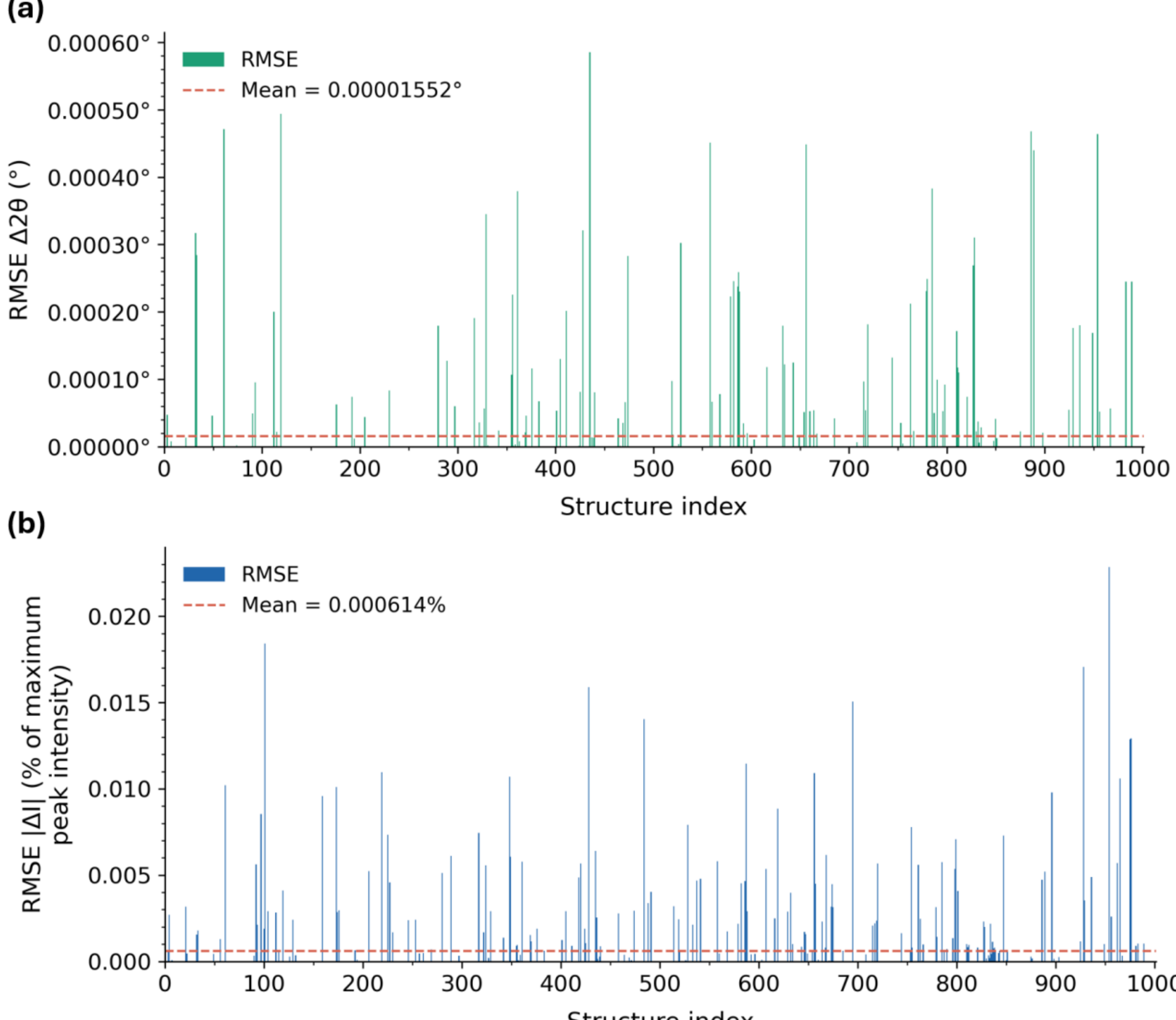

**Fig. 2:** Numerical comparison of peak positions (a) and intensities (b) between XRD-Rust and original pymatgen's XRD calculator across 1 000 COD structures, each having between 500 to 1 500 atoms (total of 8 638 808 reflections, 2θ within 2 – 60 degrees, Mo radiation).

## 4. Performance

The performance of the Rust-accelerated XRD calculator was benchmarked against the original pymatgen implementation using two crystal structure databases: MC3D (33 142 structures, as of 29/12/2025) and COD (521 900 structures, revision 297631). Powder XRD patterns were calculated over a 2 – 60° 2θ range using Mo Kα radiation. Four execution configurations of the Rust-accelerated implementation were evaluated: serial execution without SIMD (reflecting primarily the performance gain from transitioning from Python to Rust), serial execution with SIMD vectorization, 8-thread parallel execution using the Rayon library without SIMD, and 8-thread parallel execution with SIMD.

### 4.1. Performance benchmarking on the MC3D database

For serial execution with SIMD vectorization, the Rust implementation achieved a median speedup of 12.6× (MAD 2.0×) relative to the original pymatgen implementation across the MC3D dataset (**Fig. 3**). In 1 138 cases (~3.4% of structures), performance was comparable to or slightly slower than pymatgen. These cases involved only a small number of reflections, with an average runtime of ~0.14 s using the Rust-accelerated version versus ~0.08 s for the original pymatgen, making the difference practically negligible. The largest runtime reduction was observed for structure mc3d-80944, for which a diffraction pattern containing 18 948 reflections required 40.5 s with pymatgen but only 0.9 s with the Rust implementation.

To evaluate the additional contribution of SIMD vectorization and multi-threaded parallelization, the MC3D dataset was overall benchmarked across four execution configurations: serial without SIMD, serial with SIMD, 8-thread parallelization without SIMD, and 8-thread parallelization with SIMD (**Fig. 4**). The MC3D dataset consists predominantly of small structures (having between 1 and 80 atoms), for which thread initialization and SIMD setup represent a non-negligible fraction of the total calculation time, limiting the potential gain of these additional optimizations. As described above, serial execution with SIMD achieved a median speedup of 12.6× (MAD 2.0×). Compared to 10.5× (MAD 1.8×) for serial execution without SIMD (corresponding primarily to the language transition from Python to Rust), it demonstrates about an 19% additional speedup from SIMD vectorization alone. The 8-thread parallelization without SIMD yielded a speedup of 14.2× (MAD 3.7×). Combining 8-thread parallelization with SIMD provided the highest median speedup of 14.6× (MAD 3.8×), with the SIMD effect providing only a marginal ~3% improvement. The relatively small improvement of parallelization can be attributed to the absence of larger structures in MC3D (none exceeding 80 atoms), which constrains the effectiveness of reflection-level parallelism. This trend contrasts with the COD benchmark discussed in the following section, where the presence of substantially larger structures leads to a much stronger impact of parallelization. For structures with few reflections, parallelization and SIMD overhead can partially offset computational gains, whereas for structures with large numbers of reflections both threading and SIMD provide substantially greater acceleration.

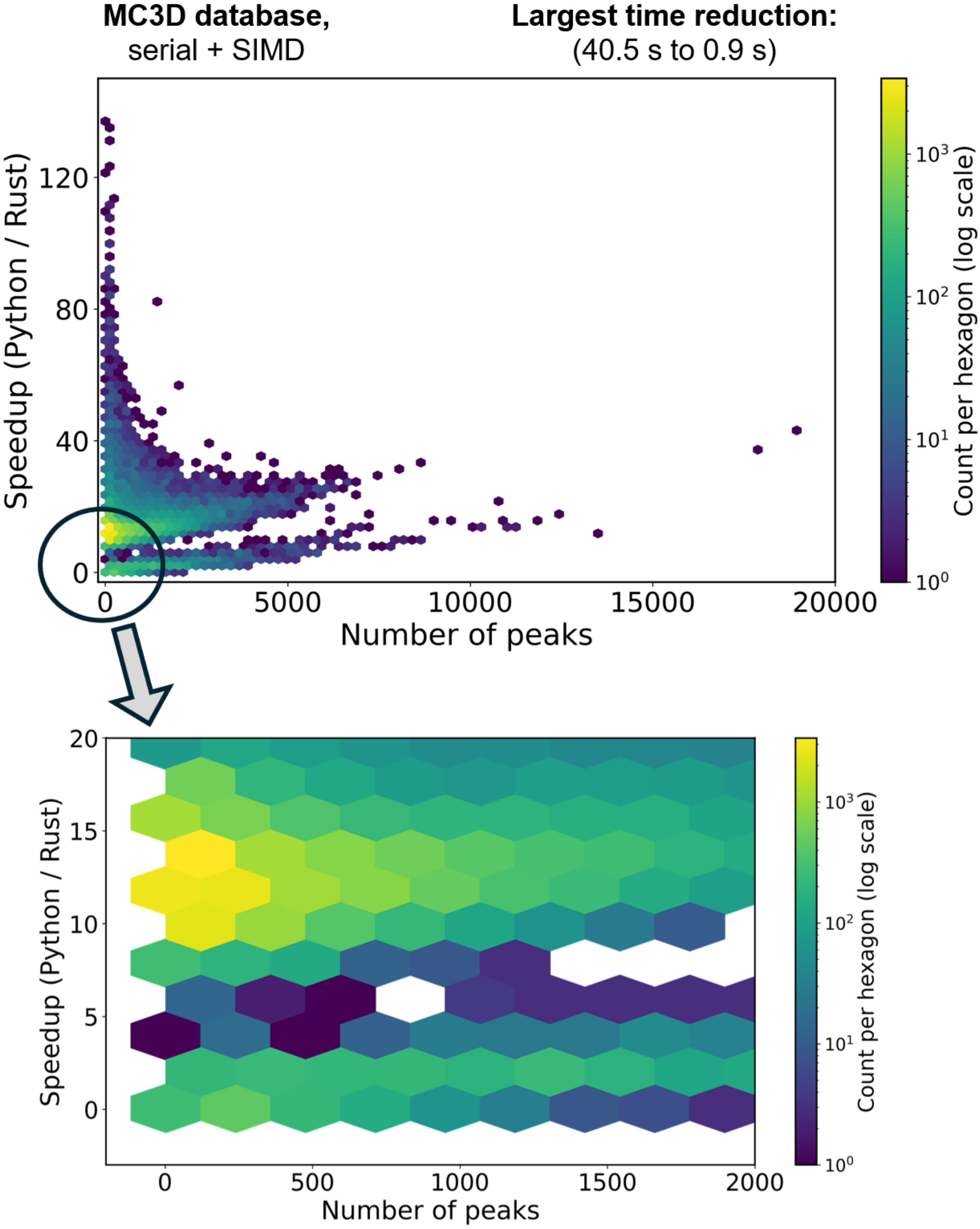

**Fig. 3:** Performance comparison between the Rust-accelerated XRD calculator with serial execution and SIMD vectorization and the original pymatgen implementation. Benchmarks were performed on 33 142 crystal structures from the MC3D database, with powder XRD patterns computed in the 2 - 60° $2\theta$ range using Mo radiation. The median speedup is 12.6× (MAD 2.0×).

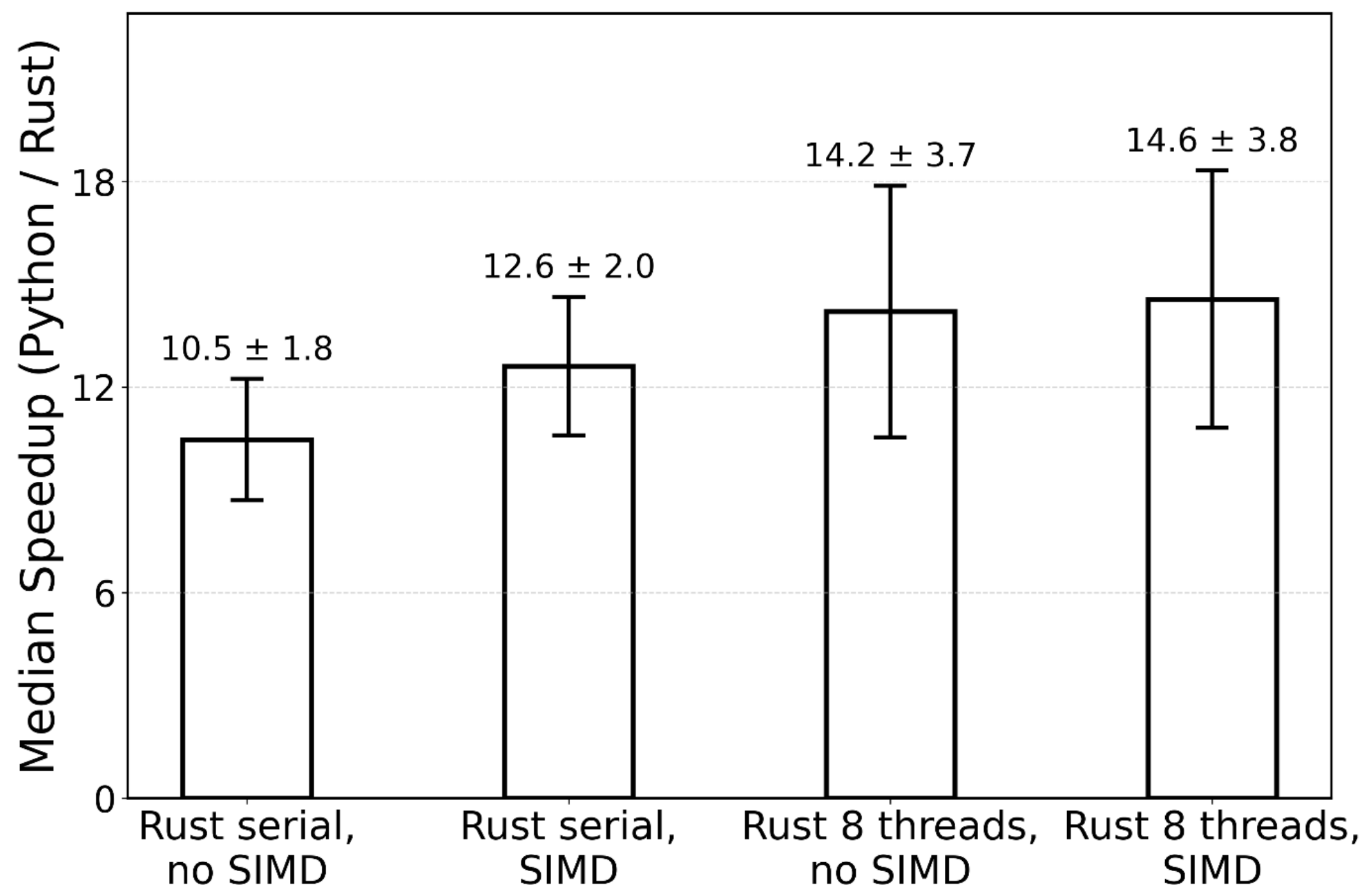

**Fig. 4:** Median speedups with ± MADs for 33 142 crystal structures from the MC3D database, comparing different execution configurations of the Rust-accelerated XRD calculator against the original pymatgen implementation.

## 4.2. Performance benchmarking on the COD database

For the COD dataset, benchmark performance is visualized with the x-axis limited to approximately 130 000 reflections and the y-axis to a 160× speedup, although a small number of structures exceeded these thresholds (**Fig. 5**). For 2 671 structures (0.5 % of structures), performance was slightly worse than or comparable to the original pymatgen implementation. Similarly to the MC3D case, this occurred for structures with a small number of reflections, where the average computation time increased from 0.2 s with pymatgen to 0.3 s with the Rust implementation. For the serial Rust implementation with SIMD, the median speedup was 10.7× (MAD 4.2×), with a maximum observed reduction in runtime from 1 437 min to 1 min (structure ID: 7201223).

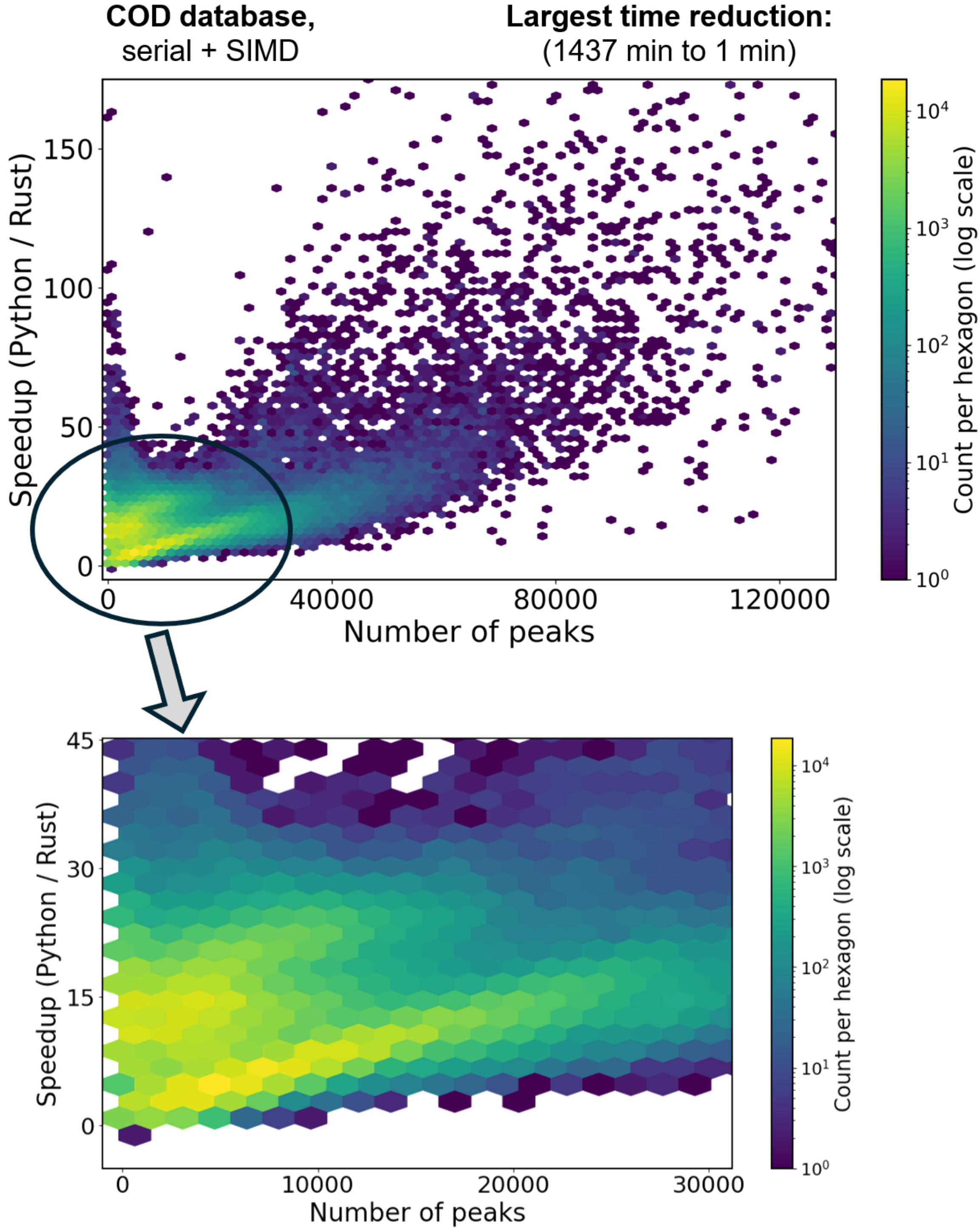

**Fig. 5:** Performance comparison between the Rust-accelerated XRD calculator with serial execution and SIMD vectorization and the original pymatgen implementation. Benchmarks were performed on 515 181 crystal structures from the COD database, with powder XRD patterns computed in the 2 - 60° 2θ range using Mo radiation. The median speedup is 10.7× (MAD 4.2×).

The maximum time reduction case, structure ID 7201223, corresponds to a large low-symmetry porous organic crystal with lattice parameters on the order of 50 Å, containing 940 atoms and yielding over 600 000 reflections in the 2 – 60° 2θ range (**Fig. 6**)[29]. The large unit cell results in a very dense reciprocal lattice, and the consequently large number of reflections combined with the atom count leads to very high number of individual structure factor evaluations. In pymatgen, these are executed within nested Python for loops where interpreter overhead dominates the total runtime at this scale. XRD-Rust largely avoids this overhead through compiled machine code, while SIMD vectorization and multi-threaded execution provide additional acceleration for structures of this kind. To verify that this result was not an outlier caused by transient benchmark effects, the calculation was repeated independently and yielded a comparable speedup.

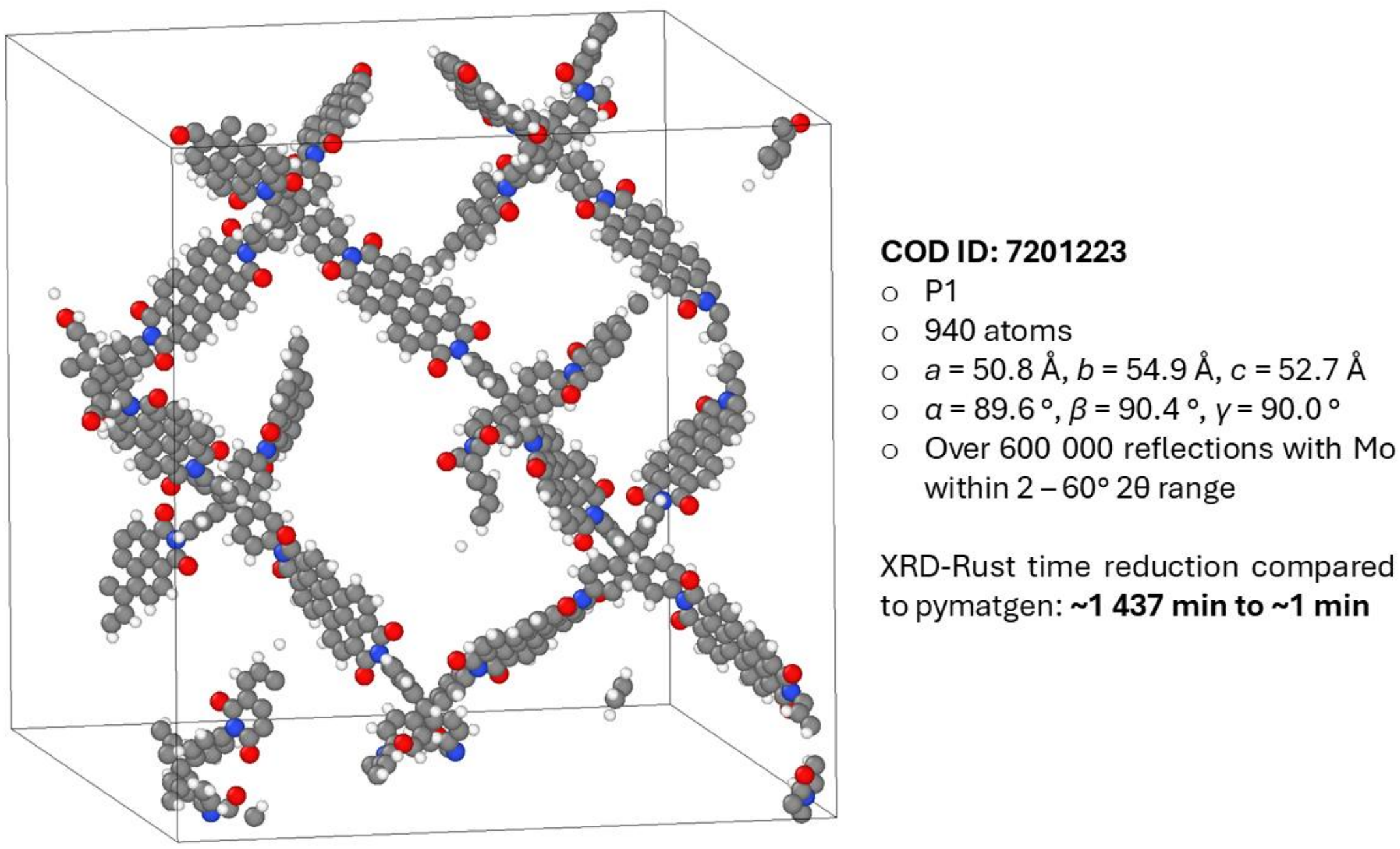

**Fig. 6:** Structure from COD database (ID-7201223) having the largest time reduction (1 437 min to 1 min) for powder diffraction calculations pattern by XRD-Rust compared to the original pymatgen implementation.

To evaluate the additional contribution of SIMD vectorization and multi-threaded parallelization, the COD dataset was further benchmarked across the same four execution configurations as described for MC3D (**Fig. 7**). Serial execution without SIMD worsened to 8.1× (MAD 3.2×), while 8-thread parallelization without SIMD yielded 18.2× (MAD 9.0×). Combining 8-thread parallelization with SIMD provided the highest median speedup of 19.5× (MAD 10.0×). SIMD vectorization contributes an additional performance gain of approximately 32% in serial execution, whereas under 8-thread parallelization its benefit is smaller, around 7%.

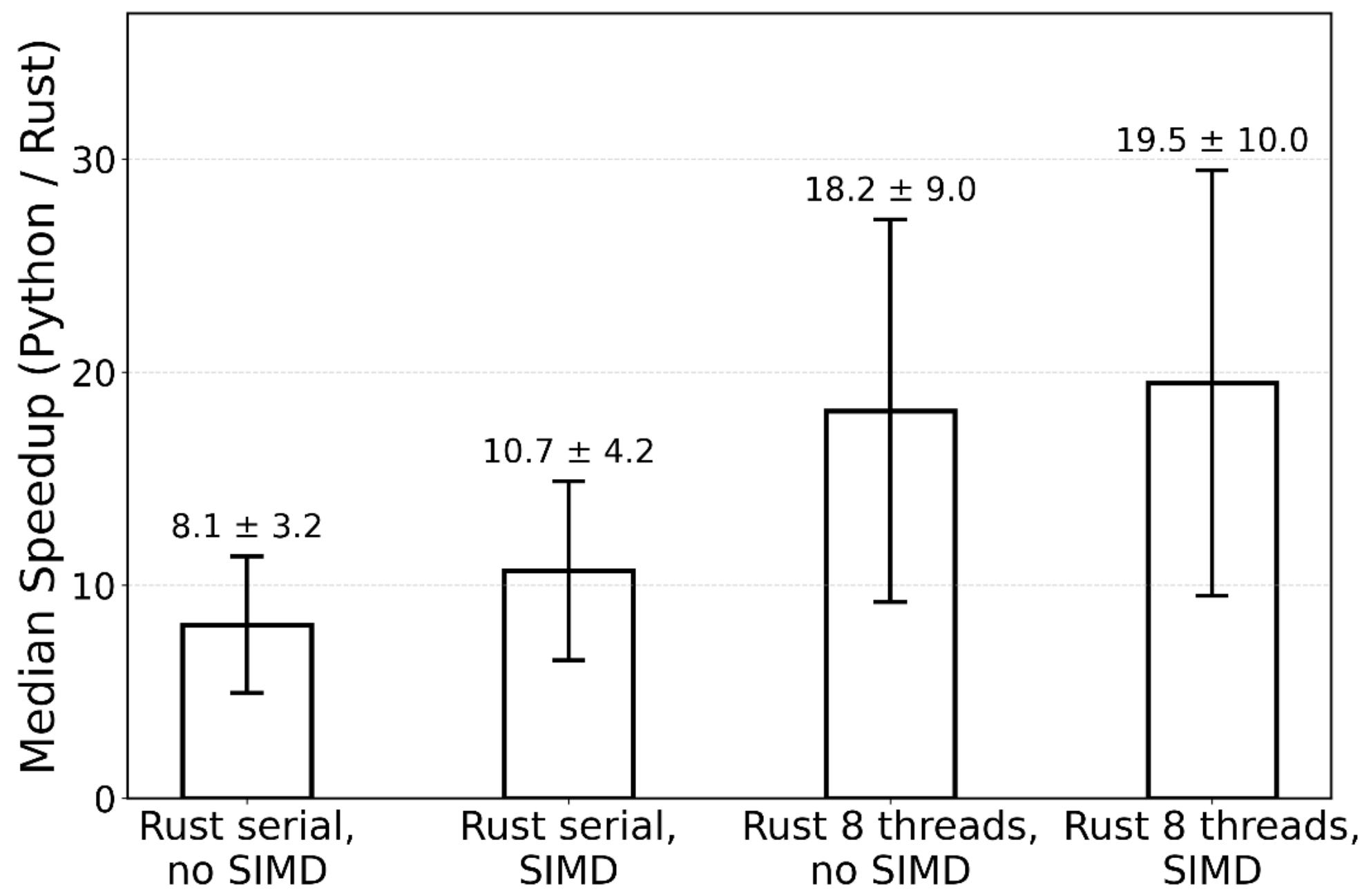

**Fig. 7:** Median speedups with ± MADs for 515 181 crystal structures from the COD database, comparing different execution configurations of the Rust-accelerated XRD calculator against the original pymatgen implementation.

To highlight the pronounced effect of thread parallelization for structures with large numbers of reflections, the speedups were grouped into four bins based on the number of reflections (**Fig. 8**): ≤10 reflections, 10 – 100 reflections, 100 – 100 000 reflections, and >100 000 reflections. For the category with the fewest reflections, parallelization yielded, as expected, lower speedups (median 10.3× with SIMD) than serial execution (15.7×), as the overhead associated with parallel execution outweighs its benefit. For structures with 10 – 100 reflections, parallelization provided a modest improvement, increasing the median speedup from 12.9× for serial execution to 15.1×. In the third bin (100 – 100 000 reflections), the serial median decreased slightly to 10.5×, whereas parallel execution further increased the median speedup to 19.8×. For structures with more than 100 000 reflections, the effect became substantial, with median speedups increasing from 118.7× for serial execution to 527.1× with parallelization. The SIMD improvement is most pronounced for the last bin with parallelization, providing further speedups about 36%.

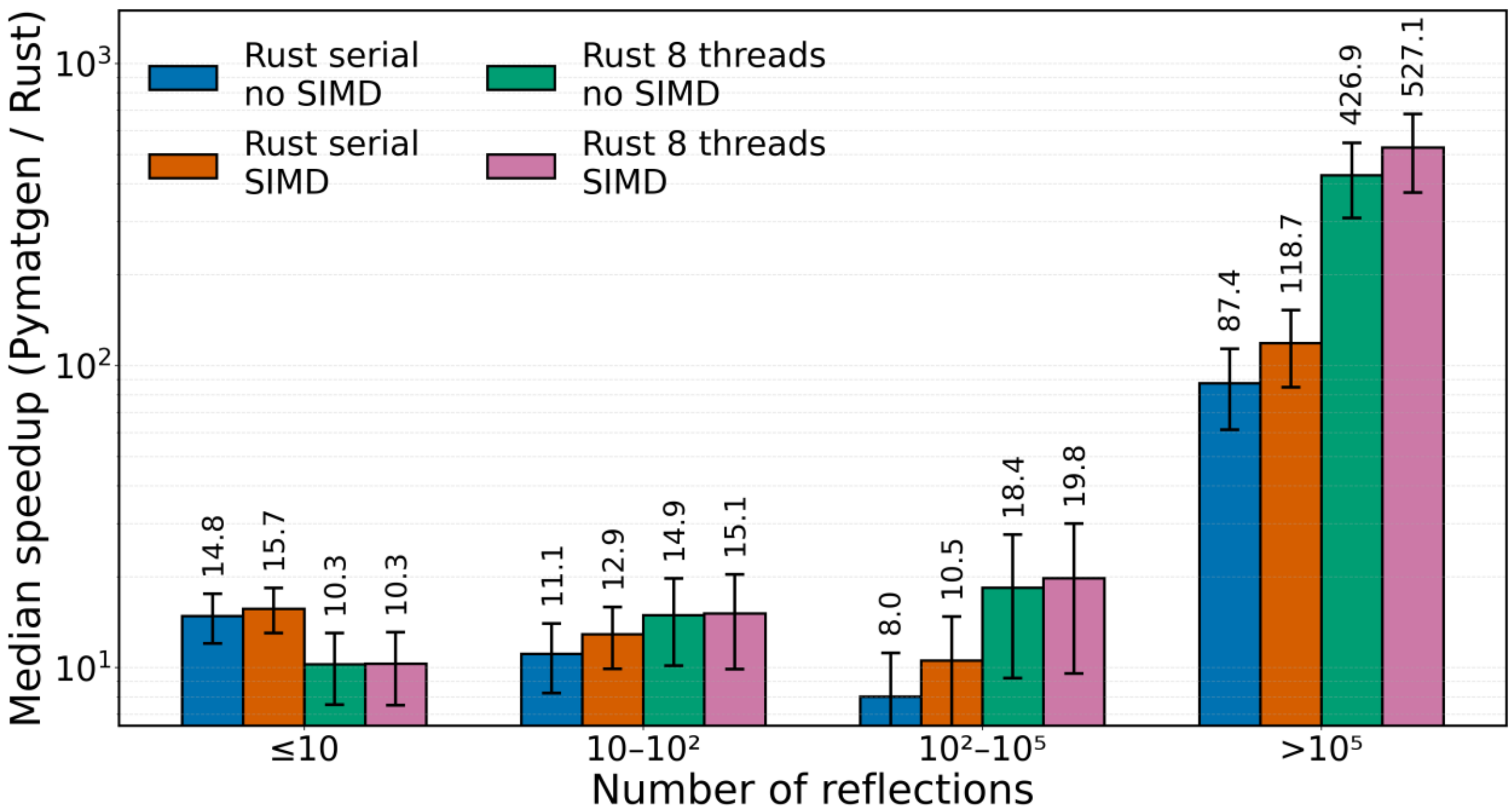

**Fig. 8:** Median speedups with ± MADs of XRD-Rust relative to the original pymatgen implementation, grouped by number of reflections, illustrating the increasing effect of thread parallelization for structures with larger number of reflections.

Overall, these benchmarks demonstrate that XRD-Rust substantially reduces the computational cost of powder XRD simulations compared with the original pymatgen implementation. The performance gains arising from thread parallelization depend on the number of reflections: for structures with few reflections, parallelization provides only modest improvement and may be partially offset by threading overhead, whereas for structures with large numbers of reflections it yields the largest speedups. SIMD vectorization provides an additional performance benefit across configurations, ranging from a few percent to few tens of percent. Together, these improvements enable more practical high-throughput powder XRD dataset generation for machine-learning workflows and improve responsiveness in interactive applications such as XRDlicious. Importantly, the Python interface and notation remain closely aligned with the original pymatgen implementation, allowing accelerated calculations to be integrated into existing workflows with minimal code modification.

## 5. Future Directions

While the present work focuses on performance optimization at the implementation level, several directions for further improvement can be considered. From an algorithmic perspective, future work may explore more efficient approaches to structure factor evaluation, including methods based on fast Fourier transforms (FFT) as well as improved reciprocal-space partitioning strategies. From a computational perspective, additional acceleration could be achieved by extending the implementation to GPU architectures, enabling improved performance for larger systems and datasets.

## 6. Conclusions

In this work, we presented XRD-Rust, a Rust-accelerated implementation of the pymatgen powder XRD calculator that substantially improves the computational performance of powder X-ray diffraction simulations while maintaining compatibility with existing pymatgen-based Python workflows. By reimplementing the computationally intensive parts of the calculation in Rust, with optional SIMD vectorization and multi-threaded execution via the Rayon library, significant speedups were achieved across large crystallographic datasets while preserving numerical fidelity with the original implementation.

Benchmarks on the MC3D database showed that the performance gain from transitioning the computational core from Python to Rust alone yielded a median speedup of 10.5× (MAD 1.8×), increasing to 12.6× (MAD 2.0×) with SIMD vectorization. Combining 8-thread parallelization with SIMD provided the highest median speedup of 14.6× (MAD 3.8×), with a maximum reduction in computation time from 40.5 s to 0.9 s. For the COD database, serial execution without SIMD yielded a median speedup of 8.1× (MAD 3.2×), increasing to 10.7× (MAD 4.2×) with SIMD vectorization, while 8-thread parallelization achieved 18.2× (MAD 9.0×) without SIMD and 19.5× (MAD 10.0×) with SIMD. A maximum runtime reduction from 1437 min to 1 min was observed for a large low-symmetry porous crystal containing more than 600 000 reflections. The impact of parallelization became particularly pronounced for structures with very large numbers of reflections, where median speedups for structures with more than 100 000 reflections increased from 118.7× in serial execution to 527.1× with 8-thread parallelization (both with SIMD).

These results demonstrate that XRD-Rust substantially reduces the computational cost of powder XRD simulations compared to the original pymatgen implementation, enabling more practical high-throughput dataset generation for machine-learning-driven materials science, while also improving responsiveness in interactive applications such as XRDlicious and keeping the original pymatgen notation.

## 7. Code Availability

The code is publicly available on GitHub at github.com/bracerino/xrd-rust and can be installed from PyPi (pypi.org/project/xrd-rust) using '*pip install xrd-rust*'. The XRD-Rust package was employed into the interactive online application (xrdlicious.com) to speed up calculations of powder XRD patterns.

## 8. Acknowledgements

This work was supported by the Grant Agency of the Czech Technical University in Prague [grant No. SGS24/121/OHK2/3T/12].